\documentclass[twocolumn,english,prl]{revtex4}
\usepackage[T1]{fontenc}
\usepackage[latin9]{inputenc}
\usepackage{color}
\usepackage{amsmath}
\usepackage{graphicx}
\usepackage{amssymb}

\makeatletter

\providecommand{\tabularnewline}{\\}

\@ifundefined{textcolor}{}
{%
 \definecolor{BLACK}{gray}{0}
 \definecolor{WHITE}{gray}{1}
 \definecolor{RED}{rgb}{1,0,0}
 \definecolor{GREEN}{rgb}{0,1,0}
 \definecolor{BLUE}{rgb}{0,0,1}
 \definecolor{CYAN}{cmyk}{1,0,0,0}
 \definecolor{MAGENTA}{cmyk}{0,1,0,0}
 \definecolor{YELLOW}{cmyk}{0,0,1,0}
 }

\makeatother

\usepackage{babel}

\begin{document}

\title{Bridging the gap by shaking superfluid matter}

\author{Mark G. Alford,$^{1}$ Sanjay Reddy,$^{2}$ and Kai Schwenzer$^{1}$}

\affiliation{$^{1}$Department of Physics, Washington University, St. Louis, MO
63130, USA}

\affiliation{$^{2}$Institute for Nuclear Theory, University of Washington, Seattle,
WA 98195-1550, USA}
\begin{abstract}
In cold compact stars, Cooper pairing between fermions in dense matter
leads to the formation of a gap in their excitation spectrum and typically
exponentially suppresses transport properties. However, we show here
that weak Urca reactions become strongly enhanced and approach their
ungapped level when the star undergoes density oscillations of sufficiently
large amplitude. We study both the neutrino emissivity and the bulk
viscosity due to direct Urca processes in hadronic, hyperonic and
quark matter and discuss different superfluid and superconducting
pairing patterns.
\end{abstract}
\maketitle
The dense matter in a compact star has certain transport properties,
such as bulk viscosity and neutrino emissivity, that are dominated
by beta (weak interaction) equilibration processes. However, this
equilibration rate is expected to be exponentially suppressed as the
star cools below the critical temperature for superconductivity/superfluidity,
and Cooper pairing produces a gap in the relevant fermion excitation
spectrum \cite{Yakovlev:2000jp,Haensel:2000vz}. This suppression
is relevant both in hadronic matter, where the critical temperature
for proton superconductivity and neutron superfluidity is in the range
$T_{c}=10^{8}-10^{10}\,\mathrm{K}$ \cite{Dean:2002zx} and in exotic
phases such as quark matter, where critical temperatures as high as
$T_{c}\simeq3-5\times10^{11}\,\mathrm{K}$ if flavor antisymmetric
pairing channels are favored , and as low as $T_{c}\simeq10^{7}\,\mathrm{K}$
when flavor singlet pairing is favored \cite{Alford:2007xm}.

In this letter we show that the exponential suppression operates only
at small amplitude and may be overcome in realistic situations.  It
is already known that beta-equilibration rates in normal matter are
enhanced by suprathermal processes at high amplitude when equilibration
is driven by higher order terms in the chemical potential imbalance
\cite{Madsen:1992sx,Reisenegger:1994be,Alford:2010gw}. Here we report
on a similar but more dramatic phenomenon in superfluid/superconducting
phases of matter, arising from a threshold-like behavior with a rapid
increase in available phase space when the typical energy in the equilibration
processes approaches the gap.

The mechanism we discuss is generic and can be expected to operate
in all situations where large perturbations drive the system out of
beta-equilibrium. One possible application are unstable oscillations
of rotating compact stars, like f- or r-modes \cite{Stergioulas:2003yp}.
These modes are unstable due to gravitational wave emission and their
amplitude grows exponentially until they are saturated by non-linear
coupling to damped daughter modes \cite{Arras:2002dw} or damped by
the supra-thermal bulk viscosity \cite{Alford:2011pi}. Note, that
even at amplitudes well below the level where bulk viscosity alone
would saturate such modes, suprathermal effects can noticeably affect
the thermal evolution of the star, via viscous heating and suprathermally
enhanced neutrino emissivity \cite{Reisenegger:1994be}. Another class
of scenarios stems from large amplitude oscillations caused by singular
events, like star quakes \cite{Franco:2000ApJ} or tidal forces preceding
neutron star mergers \cite{Tsang:2011ad}. In this case the suprathermal
bulk viscosity should control the amplitude of the oscillation and
correspondingly the emitted radiation.

\begin{figure}
\includegraphics[scale=0.38]{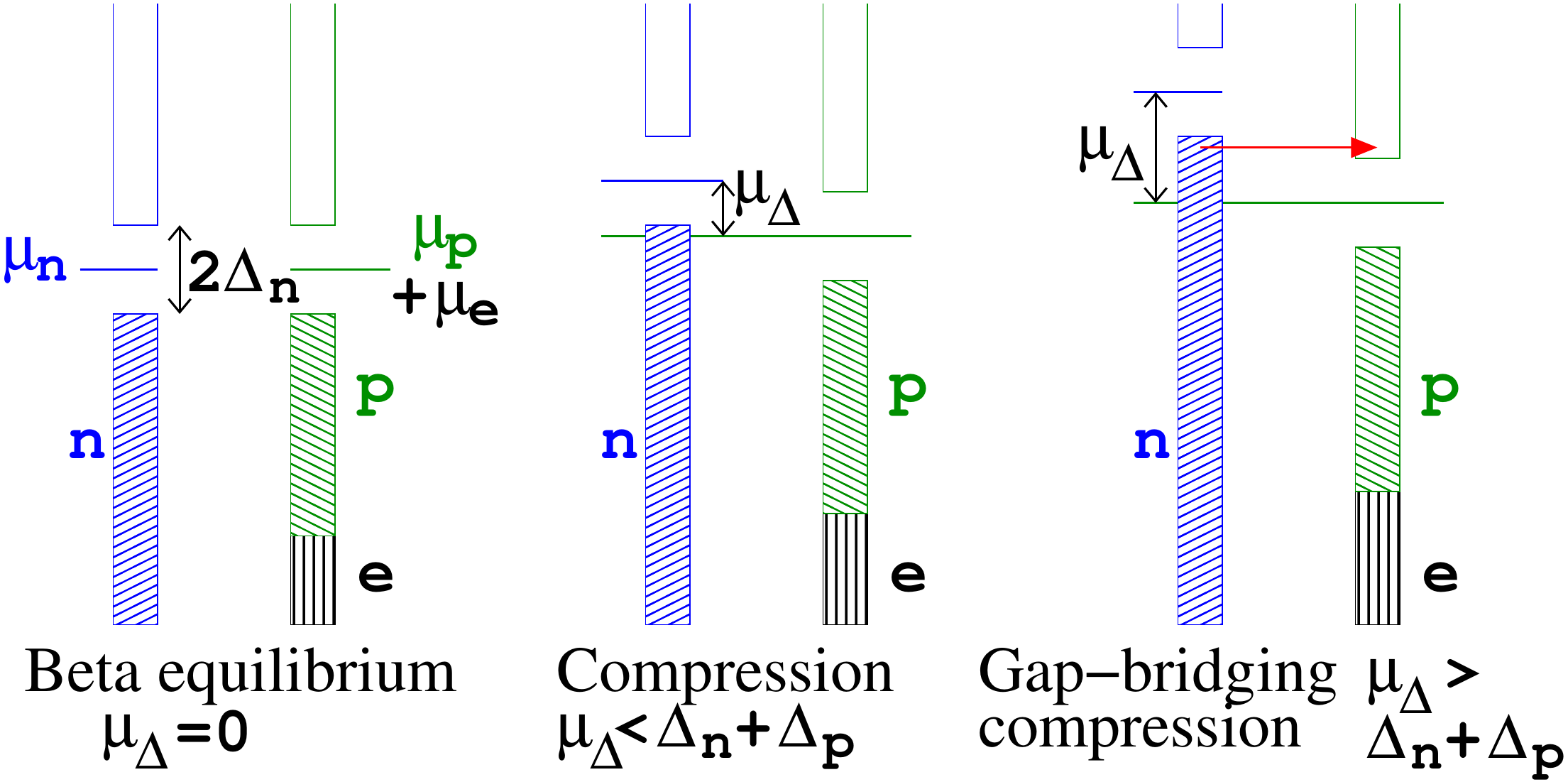}

\caption{\label{fig:fermi-seas}Schematic illustration of the opening of phase
space for Urca interactions in gapped nuclear matter when the deviation
from beta equilibrium $\mu_{\Delta}\!=\!\mu_{n}\!-\!\mu_{p}\!-\!\mu_{e}$
becomes sufficiently large relative to the pairing gaps $\Delta_{n}$,$\Delta_{p}$.}

\end{figure}

Here we will study the effect of large amplitude oscillations in the
case of various \textit{gapped} phases of dense matter. When, as in
nuclear matter, Cooper pairs contain two particles of the same flavor,
density oscillations cannot lead to pair breaking. (Thermal pair-breaking
has been used \cite{Page:2010aw,Shternin:2010qi} to explain the rapid
cooling observed in Cas A \cite{Heinke:2010cr}.) Nevertheless, oscillations
displace the gapped Fermi seas, and at sufficiently large amplitude
can \emph{bridge the gap}, opening phase space for beta equilibration
processes as illustrated in fig.~\ref{fig:fermi-seas}. As we will
show, this suprathermal effect, which begins at amplitudes far below
the level where non-linear hydrodynamic effects, of higher order in
the fluid velocity, would arise \cite{Madsen:1992sx,Alford:2010gw},
causes a huge increase in the corresponding rates, effectively bringing
them up to their ungapped level.

Beta equilibration processes have a finite time scale, so a harmonic
local density fluctuation with amplitude $\Delta n$ around the equilibrium
value $\bar{n}$ induces an oscillation in the displacement $\mu_{\Delta}$
of the chemical potentials of the degenerate particles from their
equilibrium. This oscillation leads both to heat generation (via dissipation
due to bulk viscosity) and heat loss (from enhanced neutrino emission).
Which of the two effects dominates depends among other factors on
the amplitude of the oscillation \cite{Reisenegger:1994be}. In \cite{Alford:2010gw}
we found that over the entire range of physically reasonable density
amplitudes the chemical potential oscillation is given by the linear,
harmonic relation

\begin{equation}
\mu_{\Delta}\!\left(t\right)=\hat{\mu}_{\Delta}\sin\!\left(\omega t\right)\:,\quad\hat{\mu}_{\Delta}=C\frac{\Delta n}{\bar{n}}\:,\quad C\equiv\left.\bar{n}\frac{\partial\mu_{\Delta}}{\partial n}\right|_{x}\label{eq:chem-pot-amp}\end{equation}
where $C$ is a susceptibility characterizing the particular form
of dense matter.

We are interested in the direct {}``Urca'' processes $d\rightarrow u+l+\bar{\nu}_{l}$
and $u+l\rightarrow d+\nu_{l}$ mediated by charged current interactions,
which occur in hadronic matter at high densities where proton fraction
$x\gtrsim10\%$ \cite{Lattimer:1991ib}. To study all relevant cases
within a unified framework we use a notation where $d$ represents
either an elementary negatively charged quark (down or strange) or
a hadron that contains that quark. Similarly $u$ stands either for
an elementary up quark or a hadron containing it; $l$ represents
a charged lepton (an electron, or in dense neutron matter a muon)
and $\nu_{l}$ is the corresponding neutrino. We specialize to Fermi
liquid theory, where the general dispersion relation of the matter
fields is\begin{equation}
\left(E_{i}-\mu_{i}\right)^{2}=v_{Fi}^{2}\left(p_{i}-p_{Fi}\right)^{2}+\Delta_{i}^{2}\end{equation}
where $i$ labels the matter species and $\Delta_{i}$ represents
a gap in the particle spectrum arising from superfluidity or (color)-superconductivity.
Non-Fermi liquid effects should not play a role in a gapped system
\cite{Schafer:2004jp}. We neglect dependence of the gap on energy
and momentum, but we include temperature and density dependence $\Delta_{i}=\Delta_{i}\left(T,\mu\right)$.
The leptons can to very good approximation be described by a free
dispersion relation $E_{l}^{2}\!=\! p_{l}^{2}\!+\! m_{l}^{2}$, $E_{\nu}\!=\! p_{\nu}$.
Beta equilibrium with respect to the Urca processes enforces $\mu_{\Delta}\!\equiv\!\mu_{d}\!-\!\mu_{u}\!-\!\mu_{l}\!=\!0$.
Since compact stars contain ultradegenerate matter with $T\ll\mu$
we can restrict the analysis to leading order in $T/\mu$ where we
find the general results for the net rate and the emissivity

\begin{align}
\Gamma_{dU}^{\left(\leftrightarrow\right)} & \approx\frac{17G^{2}}{120\pi}D\Theta T^{4}\mu_{\Delta}R_{\Gamma}\negmedspace\left(\frac{\mu_{\Delta}}{T},\frac{\Delta_{d}}{T},\frac{\Delta_{u}}{T}\right)\label{eq:Gamma-gen}\\
\epsilon_{dU} & \approx\frac{457\pi G^{2}}{5040}D\Theta T^{6}R_{\varepsilon}\negmedspace\left(\frac{\mu_{\Delta}}{T},\frac{\Delta_{d}}{T},\frac{\Delta_{u}}{T}\right)\label{eq:epsilon-gen}\end{align}
In these expressions $G$ is the effective coupling of the fields
to the weak current, $D$ is a function depending on the density and
the equation of state

\begin{equation}
D=\frac{p_{Fd}p_{Fu}}{\mu_{u}v_{Fd}v_{Fu}}\left(\mu_{d}^{2}\!-\! p_{Fd}^{2}\!-\!\mu_{u}^{2}\!+\! p_{Fu}^{2}\!-\! m_{l}^{2}\right)\end{equation}
which is identical for both quantities, and the parenthesis vanishes
for a free massless dispersion relation. Aside from the neglected
temperature corrections any modification of the dispersion relations
due to interactions or masses opens phase space and yields a non-zero
result. Eqs.~(\ref{eq:Gamma-gen}) and (\ref{eq:epsilon-gen}) also
contain a threshold function $\Theta\!\approx\!\theta\!\left(p_{Fu}\!+\! p_{Fl}\!-\! p_{Fd}\right)$,
a characteristic temperature and amplitude dependence, and a \emph{modification
function} $R$ given for singlet gaps by the dimensionless integrals

\begin{align}
 & R_{i}\!\left(\frac{\mu_{\Delta}}{T},\frac{\Delta_{d}}{T},\frac{\Delta_{u}}{T}\right)=N_{i}\int_{0}^{\infty}\! dx_{\nu}x_{\nu}^{2+\lambda_{i}}\nonumber \\
 & \cdot\left(\int_{-\infty}^{-\frac{\Delta_{d}}{T}}\negmedspace+\!\int_{\frac{\Delta_{d}}{T}}^{\infty}\right)\frac{dx_{d}\left|x_{d}\right|}{\sqrt{x_{d}^{2}-\frac{\Delta_{d}^{2}}{T^{2}}}}\left(\int_{-\infty}^{-\frac{\Delta_{u}}{T}}\negmedspace+\!\int_{\frac{\Delta_{u}}{T}}^{\infty}\right)\frac{dx_{u}\left|x_{u}\right|}{\sqrt{x_{u}^{2}-\frac{\Delta_{u}^{2}}{T^{2}}}}\nonumber \\
 & \cdot\tilde{n}\!\left(x_{d}\right)\tilde{n}\!\left(-\! x_{u}\right)\left(\tilde{n}\!\left(\! x_{\nu}\!+\! x_{u}\!-\! x_{d}\!-\!\frac{\mu_{\Delta}}{T}\!\right)\right.\nonumber \\
 & \qquad\qquad\qquad\qquad\left.-\left(-1\right)^{\lambda_{i}}\tilde{n}\!\left(\! x_{\nu}\!+\! x_{u}\!-\! x_{d}\!+\!\frac{\mu_{\Delta}}{T}\!\right)\right)\label{eq:R-gen}\end{align}
where $N_{\Gamma}\!=\!\tfrac{60}{17\pi^{4}}$, $N_{\epsilon}\!=\!\tfrac{2520}{457\pi^{6}}$,
$\lambda_{\Gamma}\!=\!0$ and $\lambda_{\epsilon}\!=\!1$. These functions
reflect the phase space available to Urca reactions. They are normalized
so that $R\!\left(0,0,0\right)\!=\!1$ and are symmetric in the gap
parameters $R\!\left(x,y,z\right)\!=\! R\!\left(x,z,y\right)$. In
equilibrium they are pure reduction factors $R\!\left(0,\cdots\right)\!\leq\!1$
that can become extremely small \cite{Yakovlev:2000jp}. However,
out of equilibrium $\mu_{\Delta}\!\ne\!0$ these modification functions
exhibit a suprathermal enhancement for $\mu_{\Delta}/T\!>\!1$ \cite{Madsen:1992sx,Reisenegger:1994be,Alford:2010gw}
and can then greatly exceed one. They depend on the equation of state
only via dimensionless ratios and are the same for all direct Urca
processes \cite{Yakovlev:2000jp}. In the ungapped case they have
an analytic polynomial form \cite{Reisenegger:1994be}. 

The oscillation period of a compact star is much smaller than its
evolution time scale, so the relevant quantity is the averaged emissivity
$\bar{\epsilon}_{dU}\!\equiv\!\frac{1}{2\pi}\int_{0}^{2\pi}\! d\varphi\,\epsilon_{dU}\!\left(\mu_{\Delta}\!\left(\varphi\right)\right)$
which takes the same form eq.~(\ref{eq:epsilon-gen}) with the averaged
modification function $R_{\bar{\epsilon}}$ depending on $\hat{\mu}_{\Delta}$.
 The bulk viscosity of dense matter is \cite{Alford:2010gw}

\begin{align}
\zeta & =-\frac{C}{\pi\omega^{2}}\frac{\bar{n}_{*}}{\Delta\! n_{*}}\int_{0}^{2\pi}\negmedspace d\varphi\cos\left(\varphi\right)\int_{0}^{\varphi}\negmedspace d\varphi^{\prime}\Gamma^{\left(\leftrightarrow\right)}\!\left(\mu_{\Delta}\!\left(\varphi^{\prime}\right)\right)\label{eq:general-bulk-viscosity}\end{align}
Inserting the rate eq.~(\ref{eq:Gamma-gen}) and the low-amplitude
form of the chemical potential oscillation eq.~(\ref{eq:chem-pot-amp})
gives

\begin{align}
\zeta_{dU} & =\frac{17G^{2}}{120\pi}C^{2}D\Theta\frac{T^{4}}{\omega^{2}}R_{\zeta}\negmedspace\left(\frac{\hat{\mu}_{\Delta}}{T},\frac{\Delta_{d}}{T},\frac{\Delta_{u}}{T}\right)\label{eq:dU-viscosity}\end{align}
with an analogous modification function

\begin{align}
 & R_{\zeta}\negmedspace\left(\frac{\hat{\mu}}{T},\frac{\Delta_{d}}{T},\frac{\Delta_{u}}{T}\right)=-\frac{1}{\pi}\int_{0}^{2\pi}d\varphi\cos\!\left(\varphi\right)\int_{0}^{\varphi}d\varphi^{\prime}\sin\!\left(\varphi^{\prime}\right)\nonumber \\
 & \qquad\qquad\qquad\cdot R_{\Gamma}\negmedspace\left(\frac{\hat{\mu}}{T}\sin\!\left(\varphi^{\prime}\right),\frac{\Delta_{d}}{T},\frac{\Delta_{u}}{T}\right)\label{eq:R-viscosity}\end{align}
In hyperon and quark matter the viscosity is dominated by neutral
current rather than Urca processes, but for those we also expect a
high-amplitude enhancement \cite{Alford:2010gw}.

\begin{figure*}
\begin{minipage}[t]{0.5\textwidth}%
\includegraphics[scale=0.85]{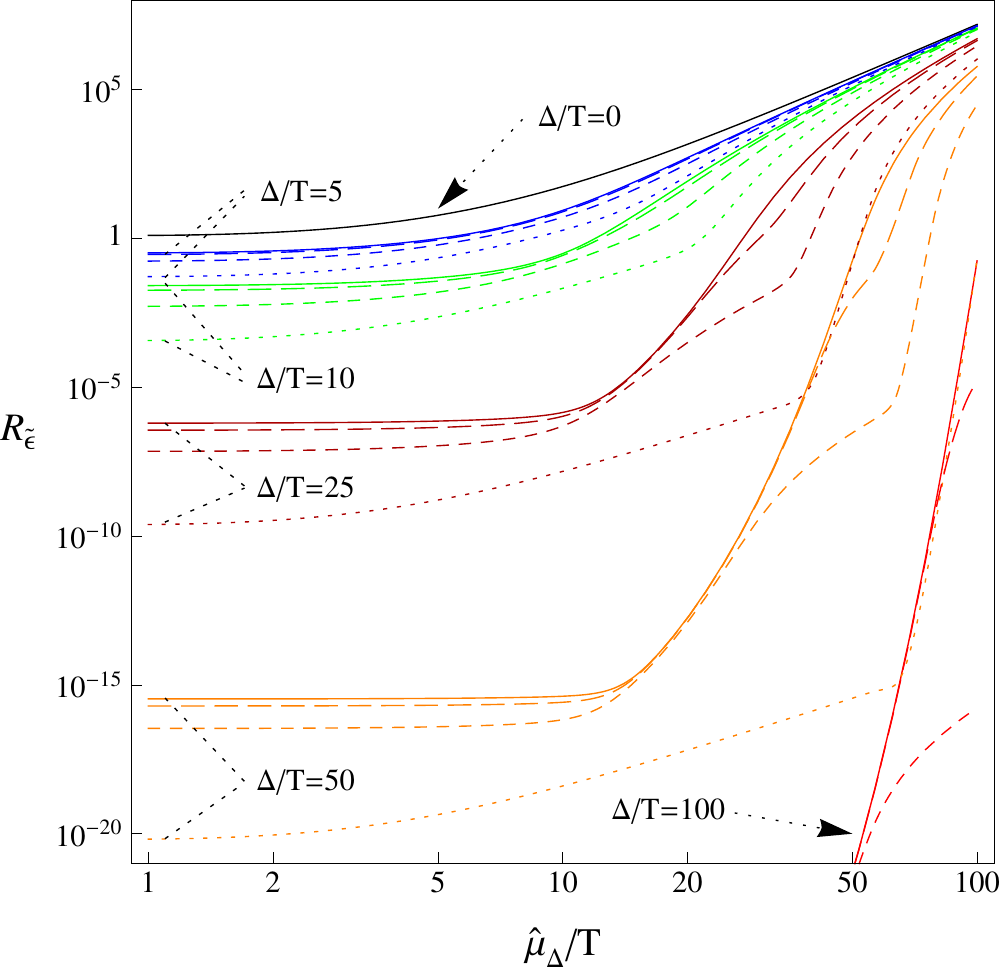}%
\end{minipage}%
\begin{minipage}[t]{0.5\textwidth}%
\includegraphics[scale=0.85]{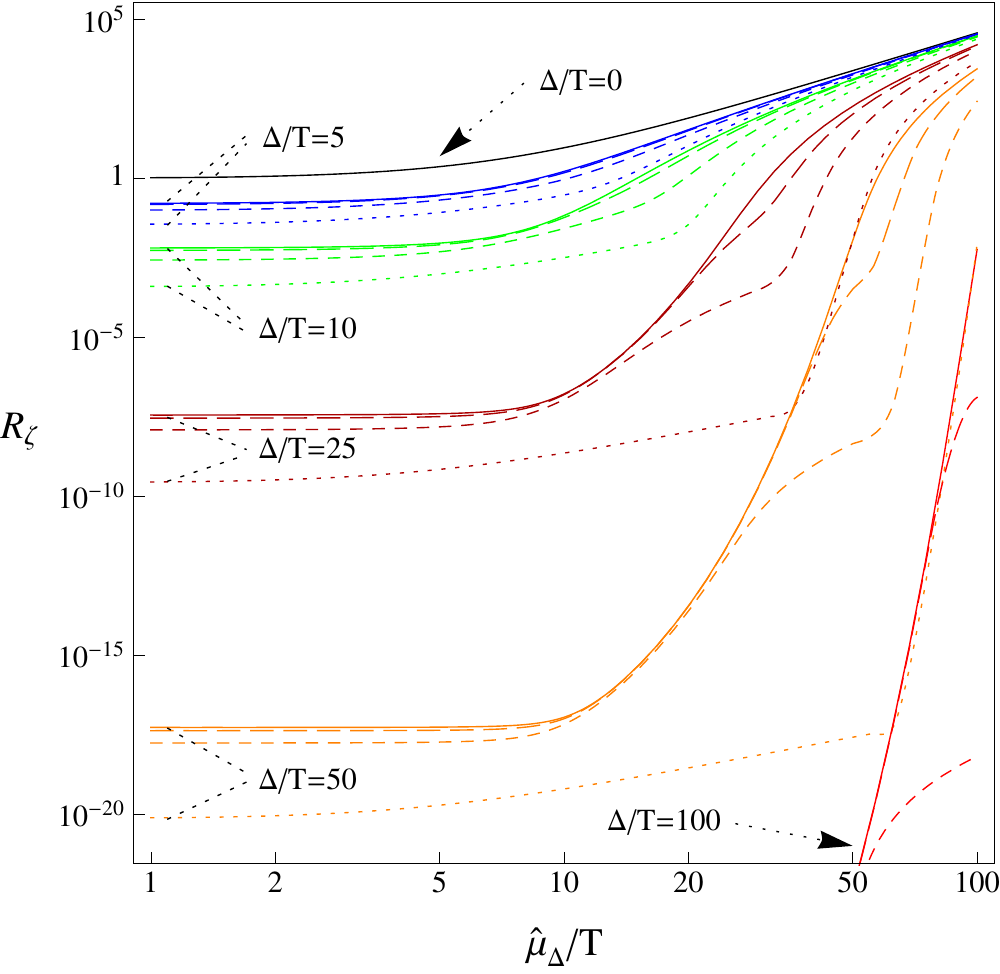}%
\end{minipage}

\caption{\label{fig:modification-functions}The modification functions $R_{i}$
eqs.~(\ref{eq:epsilon-gen}) \& (\ref{eq:dU-viscosity})  as a function
of oscillation amplitude $\hat{\mu}_{\Delta}$, for different pairing
patterns with maximum gap $\Delta$: symmetric case where only one
particle is gapped, e.g. $\Delta_{d}\!=\!\Delta$, $\Delta_{u}\!=\!0$
(solid lines); intermediate cases $\Delta_{d}\!=\!5\Delta_{u}\!=\!\Delta$
(long dashed lines); $\Delta_{d}\!=\!2\Delta_{u}\!=\!\Delta$ (short
dashed lines); symmetric case $\Delta_{d}\!=\!\Delta_{u}\!=\!\Delta$
(dotted lines). Gap ranges from $\Delta/T\!=\!0$ to $\Delta/T\!=\!100$.\emph{
Left panel:} $R_{\bar{\epsilon}}$ for the averaged neutrino emissivity.
\emph{Right panel:} $R_{\zeta}$ eq.~(\ref{eq:R-viscosity}) for
the bulk viscosity.}

\end{figure*}
\begin{figure}
\includegraphics[scale=0.85]{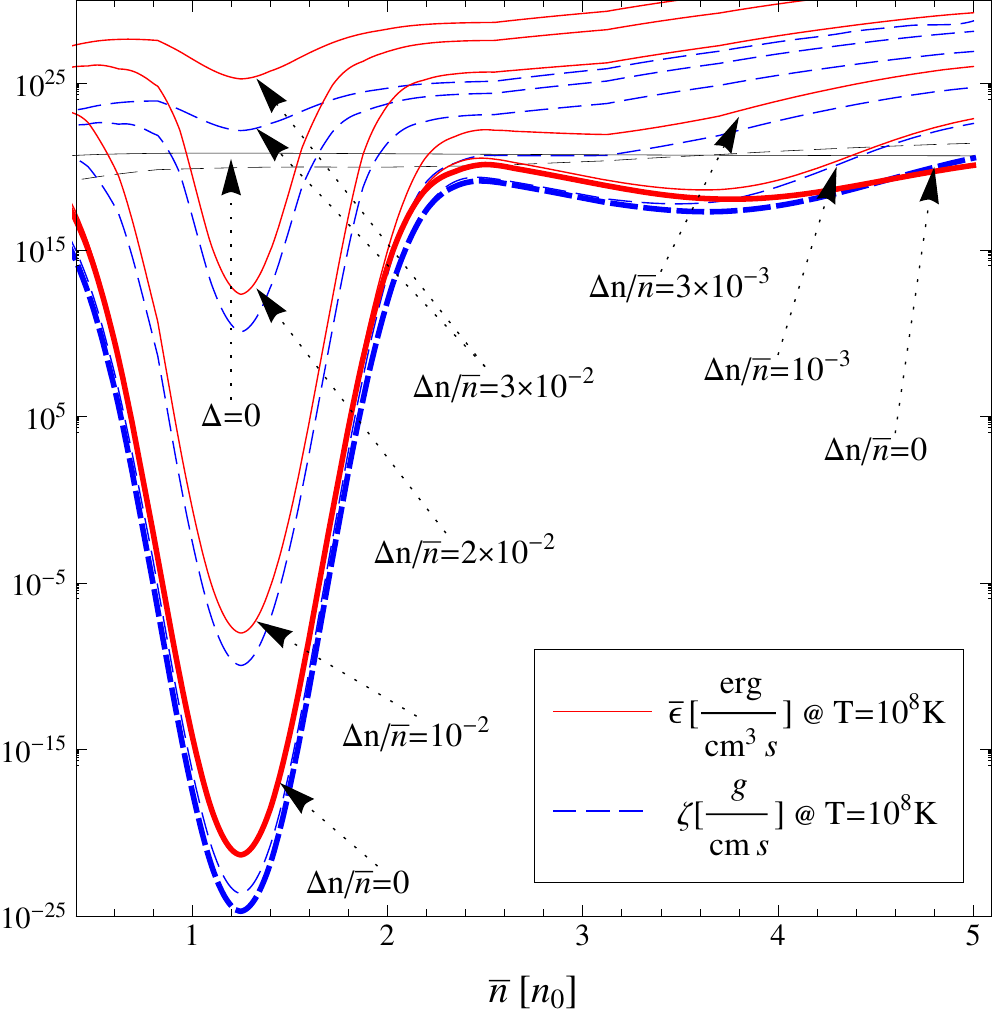}

\caption{\label{fig:hadronic-results}Enhancement via suprathermal oscillations
of the neutrino emissivity and the bulk viscosity of hadronic matter
with a $^{1}S_{0}$ proton gap $\Delta_{p0}\!=\!1\,\mathrm{MeV}$
at $n\!\approx\!1.3\, n_{0}$ and a neutron gap $\Delta_{n0}\!\approx\!0.1\,\mathrm{MeV}$
at $n\!\approx\!4\, n_{0}$.}

\end{figure}
\begin{table*}
\begin{tabular}{|c|c|c|c|c|c|c|c|c|c|c|}
\hline 
process & $G$ & $f_{V}$ & $g_{A}$ & $D$ & $\mu_{e}$ & $C$ & $\tilde{\epsilon}\left[\frac{\mathrm{ergs}}{\mathrm{cm}^{3}\mathrm{s}}\right]$ & $\tilde{\zeta}\left[\frac{\mathrm{g}}{\mathrm{cm}\,\mathrm{s}}\right]$ & $\nu_{\epsilon}$ & $\nu_{\zeta}$\tabularnewline
\hline 
$n\to pl\bar{\nu}_{l}$ & $\frac{1}{2}\sqrt{f_{V}^{2}\!+\!3g_{A}^{2}}\cos\!\theta_{C}G_{F}$ & $1$ & $1.23$ & $2m_{n}^{*}m_{p}^{*}\mu_{e}$ & $4\left(1\!-\!2x\right)S$ & $4\!\left(1\!-\!2x\right)\!\left(n\frac{\partial S}{\partial n}\!-\!\frac{S}{3}\right)$ & $6.82\cdot10^{26}$ & $7.86\cdot10^{22}$ & $\frac{2}{3}$ & $2$\tabularnewline
\cline{1-5} \cline{8-9} 
$\Lambda\to pl\bar{\nu}_{l}$ & $\frac{1}{2}\sqrt{f_{V}^{2}\!+\!3g_{A}^{2}}\sin\!\theta_{C}G_{F}$ & $-1.23$ & $0.89$ & $2m_{\Lambda}^{*}m_{p}^{*}\mu_{e}$ & or $\frac{\left(3\pi^{2}n\right)^{2/3}}{2m_{n}}$ & or $\frac{\left(3\pi^{2}n\right)^{2/3}}{6m_{n}}$ & $3.05\cdot10^{25}$ & $3.52\cdot10^{21}$ &  & \tabularnewline
\cline{1-5} \cline{8-9} 
$\Sigma^{-}\to nl\bar{\nu}_{l}$ & $\frac{1}{2}\sqrt{f_{V}^{2}\!+\!3g_{A}^{2}}\sin\!\theta_{C}G_{F}$ & $-1$ & $0.28$ & $2m_{\Sigma}^{*}m_{n}^{*}\mu_{e}$ & for a free gas & for a free gas & $1.04\cdot10^{25}$ & $1.20\cdot10^{21}$ &  & \tabularnewline
\hline 
$d\to ue\bar{\nu}_{e}$ & $\sqrt{3}\cos\!\theta_{C}G_{F}$ & $-$ & $-$ & $\frac{8\alpha_{s}}{3\pi}\mu_{q}^{2}\mu_{e}$ & $\frac{m_{s}^{2}}{4\mu_{q}}$ & $-\frac{m_{s}^{2}}{3(1-c)\mu_{q}}$ & $1.57\cdot10^{25}\alpha_{s}$ & $5.09\cdot10^{21}\alpha_{s}$ & $\frac{1}{3}$ & $-\frac{1}{3}$\tabularnewline
\cline{1-5} \cline{8-9} 
$s\to ue\bar{\nu}_{e}$ & $\sqrt{3}$$\sin\!\theta_{C}G_{F}$ & $-$ & $-$ & $m_{s}^{2}\mu_{q}$ &  & free gas: $c=0$ & $3.98\cdot10^{24}$ & $1.29\cdot10^{21}$ &  & \tabularnewline
\hline
\end{tabular}

\caption{\label{tab:Parameters}Parameters for direct Urca processes in dense
matter, with Fermi constant $G_{F}$, Cabbibo angle $\theta_{C}$,
vector and axial couplings $f_{V}$ and $g_{A}$ \cite{Yakovlev:2000jp},
proton fraction $x$, symmetry energy $S\!\left(\bar{n}\right)$ \cite{Lattimer:1991ib},
in-medium hadron masses $m^{*}\negthickspace\left(\bar{n}\right)$
and quark interaction parameter $c$ \cite{Alford:2004pf,Alford:2010gw}.
The quark expression is to leading order in $\alpha_{s}$ and all
results to leading order in $T/\mu$, $m_{l}/\mu$ and $\mu_{\Delta}/\mu_{e}$.}

\end{table*}
To see the gap-bridging enhancement of beta-equilibration rates for
oscillations of suprathermal amplitude, without specifying a particular
form of matter, we show in fig.~\ref{fig:modification-functions}
the modification functions $R_{\bar{\epsilon}}$ and $R_{\zeta}$.
We vary the maximum gap $\Delta$ and the ratio of the two gaps. For
strongly gapped particles $\Delta_{i}/T\!\gg\!1$, the larger gap
tends to dominate, e.g. $(\Delta_{d},\Delta_{u})=(2\Delta,0)$ is
more suppressed than $(\Delta,\Delta)$.  Once the amplitude $\hat{\mu}_{\Delta}$
becomes comparable to $\Delta$, the Urca reactions can bridge the
gap(s) so that the modification functions rise steeply, and quickly
reach their ungapped levels. Interestingly, for matter where the two
gaps are the same (dotted lines) this rise starts already for $\mu_{\Delta}\!\ll\!\Delta_{i}$
and eventually merges into the solid curve for a single gap $\Delta_{d}\!+\!\Delta_{u}$.
In contrast, in the asymmetric case where one gap is much larger than
the other, the effects of the smaller gap are only visible at large
amplitudes and become negligible when it is more than an order of
magnitude smaller.

Fig.~\ref{fig:modification-functions} shows that suprathermal enhancement
in gapped matter is considerably bigger than in ungapped matter (the
$\Delta/T\!=\!0$ line) \cite{Alford:2010gw}. This enhancement could
be realistically achieved in compact star oscillations: $\hat{\mu}_{\Delta}$
is related to the density amplitude $\Delta n/\bar{n}$ by the susceptibility
$C$ eq.~(\ref{eq:chem-pot-amp}). For hadronic matter with an APR
equation of state \cite{Akmal:1998cf}, $C$ increases from $C\left(n_{0}/4\right)\!\approx\!20\,\mathrm{MeV}$
to $C\left(5n_{0}\right)\!\approx\!150\,\mathrm{MeV}$ \cite{Haensel:2000vz};
for quark matter it decreases from $C\left(n_{0}\right)\!\approx\!30\,\mathrm{MeV}$
to $C\left(5n_{0}\right)\!\approx\!20\,\mathrm{MeV}$. Thus for amplitudes
$\Delta n/\bar{n}\!\sim0.01$ the amplitude $\hat{\mu}_{\Delta}$
can indeed become large enough to bridge gaps $\Delta\!\sim1\,\mathrm{MeV}$
\cite{Alford:2010gw}.

To apply these general results to explicit processes requires dispersion
relations and effective couplings for the relevant particles. Here
we consider direct Urca processes in hadronic \cite{Lattimer:1991ib,Haensel:1992zz,Haensel:2000vz},
hyperonic \cite{Prakash:1992ApJ...390L..77P,Haensel:2000vz} and quark
matter \cite{Iwamoto:1980eb,Sa'd:2007ud}. Parameter values are given
in tab. \ref{tab:Parameters}. Inserted in the general results eqs.~(\ref{eq:epsilon-gen})
and (\ref{eq:dU-viscosity}) they reproduce previous expressions in
the subthermal case $\mu_{\Delta}\!\ll\! T$. In the ideal gas approximation
our results simplify to\begin{equation}
\bar{\epsilon}=\tilde{\epsilon}\left(\frac{\bar{n}}{n_{0}}\right)^{\nu_{\epsilon}}\negthickspace T_{9}^{6}R_{\bar{\epsilon}}\;,\;\zeta=\tilde{\zeta}\left(\frac{\bar{n}}{n_{0}}\right)^{\nu_{\zeta}}\negthickspace\left(\frac{1\mathrm{kHz}}{\omega}\right)^{2}\negthickspace T_{9}^{4}R_{\zeta}\end{equation}
where $T_{9}$ is the temperature in units of $10^{9}\,\mathrm{K}$
and the values of $\tilde{\epsilon}$ and $\tilde{\zeta}$ are given
in tab.~\ref{tab:Parameters}. 

As an explicit example that takes into account the complications in
gapped interacting matter we consider APR hadronic matter \cite{Akmal:1998cf}
with the in-medium hadron masses given in \cite{Page:2004fy}. For
the density dependence of the $^{1}S_{0}$ proton gap we use a Gaussian
fit to data from \cite{Dean:2002zx}, with maximum $\Delta_{p0}\!=\!1\,\mathrm{MeV}$
centered at $n\!\approx\!1.3\, n_{0}$. For illustrative purposes
we assume the neutrons also have a $^{1}S_{0}$ gap (in reality it
is expected to be $^{3}P_{2}$) with the density dependence proposed
to explain the Cas A cooling data in \cite{Shternin:2010qi}, i.e.
with maximum $\Delta_{n0}\!=\!0.12\,\mathrm{MeV}$ at $n\!\approx\!3.7\, n_{0}$.
The gaps have a BCS temperature dependence, but this has only a modest
effect in small regions where $T\!\sim\!\Delta_{p},\Delta_{n}$. Our
results for neutrino emissivity and bulk viscosity at $T\!=\!10^{8}\,\mathrm{K}$
are shown in fig.~\ref{fig:hadronic-results}. At infinitesimal amplitudes
(thick $\Delta n/\bar{n}\!=0$ lines) there is enormous suppression
by the proton gap, and moderate suppression by the smaller neutron
gap. As the density oscillation amplitude increases, it first, at
$\Delta n/\bar{n}\!\gtrsim\!10^{-3}$, overcomes the suppression of
emissivity and viscosity due to the neutron gap, increasing them at
densities potentially relevant for direct Urca processes by many orders
of magnitude. Then, at $\Delta n/\bar{n}\!\gtrsim\!10^{-2}$, even
the large proton gap at lower density is bridged, boosting the results
by up to $10^{50}$! As in previous results for ungapped matter
\cite{Alford:2010gw}, these effects are expected to be even more
pronounced for the realistic case of modified Urca reactions \cite{Yakovlev:2000jp,Reisenegger:1994be},
that are actually present at moderate densities.

Fig. \ref{fig:hadronic-results} shows that, if the star is cold enough
that Urca processes are blocked by pairing gaps throughout the star,
density oscillations could provide the dominant contribution to those
processes, by bridging the pairing gaps, starting in the regions where
the rate-controlling gap is the smallest. The smaller this minimum
gap, the lower the amplitude required to bridge it. In hyperonic
\cite{Prakash:1992ApJ...390L..77P,Haensel:2000vz} or quark matter
\cite{Schmitt:2005wg,Wang:2010ydb} there are processes which are
only suppressed by $\Delta\!\lesssim\!0.01$ MeV, that could be bridged
by oscillations with amplitude as small as $\Delta n/\bar{n}\!\lesssim\!10^{-4}$.
\begin{acknowledgments}
We are grateful to C. Horowitz and J. Read for helpful discussions.
This work was initiated during the workshop INT-11-2b: {}``Astrophysical
Transients: Multi-messenger Probes of Nuclear Physics''. We thank
the Institute for Nuclear Theory at the University of Washington for
its hospitality and the U.S. Department of Energy for support under
contracts \#DE-FG02-91ER40628, \#DE-FG02-05ER41375, \#DE-FG02-00ER41132
and by the Topical Collaboration {}``Neutrinos and nucleosynthesis
in hot dense matter''. 
\end{acknowledgments}
\bibliographystyle{h-physrev}
\bibliography{cs}

\end{document}